\documentstyle[12pt,epsfig]{article}
\def\lbldef#1#2{\expandafter\gdef\csname #1\endcsname {#2}}

\def\href#1#2{#2}  

\begin{document}
\baselineskip=15.5pt
\pagestyle{plain}
\setcounter{page}{1}

\begin{titlepage}

\begin{flushright}
CERN-TH/2000-229\\
hep-th/0007252
\end{flushright}
\vspace{10 mm}

\begin{center}
{\Large Localization of Gravity on Dilatonic Domain Walls:\\
Addendum to ``Solitons in Brane Worlds''}

\vspace{5mm}

\end{center}

\vspace{5 mm}

\begin{center}
{\large Donam Youm\footnote{E-mail: Donam.Youm@cern.ch}}

\vspace{3mm}

Theory Division, CERN, CH-1211, Geneva 23, Switzerland

\end{center}

\vspace{1cm}

\begin{center}
{\large Abstract}
\end{center}

\noindent

We supplement the discussion on localization of gravity on dilatonic 
domain walls in ``Solitons in Brane Worlds'' (Nucl. Phys. {\bf B576}, 106, 
hep-th/9911218) by giving unified and coherent discussion which combines 
the result of this paper and the expanded results in the author's subsequent 
papers in an attempt to avoid misleading readers.  We also discuss the 
possible string theory embeddings of the Randall-Sundrum type brane world 
scenarios through non-dilatonic and dilatonic domain walls, which 
straightforwardly follows from the author's previous works but was not 
elaborated explicitly.   

\vspace{1cm}
\begin{flushleft}
CERN-TH/2000-229\\
July, 2000
\end{flushleft}
\end{titlepage}
\newpage

In Ref. \cite{youm1}, we argued that dilatonic domain walls with $\Delta<-2$ 
(see just below Eq. (\ref{valofk}) for definition of $\Delta$) localize 
gravity, since the potential term in the Schr\"odinger-type equation satisfied 
by the metric fluctuation is volcano-shaped just like that of the 
Randall-Sundrum (RS) domain wall \cite{rs1,rs2}.  In the subsequent paper 
\cite{youm2}, we found out that gravity can be localized around the positive 
tension domain wall with $\Delta<0$, since the effective gravitational 
constant in one lower dimensions is nonzero for such case.  Then, later 
it is realized \cite{youm3} that the graviton can be localized on the 
dilatonic domain wall for any values of $\Delta$ as long as the tension of 
the domain wall is positive, since for such case the normalizable 
Kaluza-Klein (KK) zero mode exists and the effective gravitational constant 
in one lower dimensions is nonzero.  We feel that these different (but 
expanded) results scattered around our previous works may cause confusion 
among readers.  Furthermore, perhaps to worsen the confusion, in Ref. 
\cite{pl}, which later reproduces these our previous papers, it is argued 
that gravity can be trapped on the dilatonic domain walls only for the 
$\Delta\leq -2$ case, since only for such case normalizable graviton KK mode 
exists.  It is a purpose of this brief note to gather the results on the 
localization of gravity on the dilatonic domain wall, which are scattered 
around our different previous papers, to give unified and coherent discussion 
in order to prevent misleading readers, and hopefully to clarify the 
contradicting result of Ref. \cite{pl}.  We also elaborate on the possible 
embeddings of the RS type brane world scenario into string theories through 
dilatonic and non-dilatonic domain walls, which follows straightforwardly 
from our previous works but was not elaborated explicitly.  

The action for the dilatonic domain wall with an arbitrary dilaton 
coupling parameter $a$ in arbitrary spacetime dimensions $D$ is
\begin{equation}
S_{\rm bulk}={1\over{2\kappa^2_D}}\int d^Dx\sqrt{-G}\left[{\cal R}_G
-{4\over{D-2}}\partial_M\phi\partial^M\phi+e^{-2a\phi}\Lambda\right].
\label{dildwact}
\end{equation}
The worldvolume action for the dilatonic domain wall with energy density 
(or tension) $\sigma_{DW}$ is
\begin{equation}
S_{\rm DW}=-\int d^{D-1}x\sqrt{-\gamma}\sigma_{DW}e^{-a\phi},
\label{dwwvact}
\end{equation}
where $\gamma$ is the determinant of the induced metric $\gamma_{\mu\nu}=
\partial_{\mu}X^M\partial_{\nu}X^NG_{MN}$ on the domain wall worldvolume.  

The dilatonic domain wall solution in the standard warp factor form 
is \cite{youm1,youm3}:
\begin{eqnarray}
& &G_{MN}dx^Mdx^N={\cal W}\left[-dt^2+dx^2_1+\cdots+dx^2_{D-2}\right]+dy^2, 
\cr
& &\phi={1\over a}\ln(1+Ky),\ \ \ \ \ 
{\cal W}=(1+Ky)^{8\over{(D-2)^2a^2}},
\label{warpdw}
\end{eqnarray}
where the constant $K$ can take following positive or negative value 
independently on both sides of the domain wall:
\begin{equation}
K=\pm{{(D-2)a^2}\over 2}\sqrt{{{D-2}\over{4(D-1)-a^2(D-2)^2}}\Lambda}
=\pm{{(D-2)a^2}\over 2}\sqrt{-{\Lambda\over{2\Delta}}},
\label{valofk}
\end{equation}
where $\Delta\equiv (D-2)a^2/2-2(D-1)/(D-2)$.  From this expression for $K$, 
one can see that the domain solution of the form (\ref{warpdw}) exists only 
for $\Delta<0$ [$\Delta>0$] when $\Lambda>0$ [$\Lambda<0$].  The boundary 
condition at $y=0$ fixes the domain wall tension $\sigma_{DW}$ 
to be the following fine-tuned form \cite{youm3}:
\begin{equation}
\sigma_{DW}={1\over{\kappa^2_D}}{4\over{(D-2)a^2}}(K_--K_+),
\label{brntnsn}
\end{equation}
where $K_-$ [$K_+$] denotes the value of $K$ at $y<0$ [$y>0$].  So, 
$\sigma_{DW}$ is positive [negative] when $K_->0$ and $K_+<0$ [$K_-<0$ and 
$K_+>0$], in which case there are naked singularities at finite nonzero $y$ 
[no singularity at $y\neq 0$].  And $\sigma_{DW}$ is zero when $K_+$ and 
$K_-$ have the same sign.   The $(D-1)$-dimensional gravitational constant  
has non-zero value given by $\kappa^2_{D-1}={{\Delta+4}\over 2}
\sqrt{-{\Lambda\over{2\Delta}}}\kappa^2_D$ when $\sigma_{DW}>0$; otherwise, 
$\kappa^2_{D-1}=0$ \cite{youm2,youm3}.  So, a necessary condition for 
localizing gravity on the domain wall is $\sigma_{DW}>0$, which is possible 
for {\it any values of} $\Delta$.

By redefining the transverse coordinate of the domain wall, one can put the 
solution into the following conformally flat form \cite{youm1,youm3}:
\begin{eqnarray}
& &G_{MN}dx^Mdx^N={\cal C}\left[-dt^2+dx^2_1+\cdots+dx^2_{D-2}+dz^2\right],
\cr
& &\phi={{(D-2)a}\over{2(\Delta+2)}}\ln(1+\bar{K}z),\ \ \ \ \ \ 
{\cal C}=(1+\bar{K}z)^{4\over{(D-2)(\Delta+2)}},
\label{cfdwsol}
\end{eqnarray}
where the constant $\bar{K}$ is given by 
\begin{equation}
\bar{K}=\eta{{(D-2)^2a^2-4}\over{(D-2)^2a^2}}K=\pm\eta(\Delta+2)
\sqrt{-{\Lambda\over{2\Delta}}},
\label{kbar}
\end{equation}
where $\pm$ is the same as $\pm$ in Eq. (\ref{valofk}) and the sign 
ambiguity $\eta=\pm$ resulting from the coordinate transformation to the 
conformally flat metric can be fixed to be  $\eta=+$ by demanding that 
$\kappa^2_{D-1}$ should be calculated to take the same values as above.  
From now on, we consider only the ${\bf Z}_2$-symmetric solution with 
$\sigma_{DW}>0$, for which $\kappa_{D-1}\neq 0$.  So, in the remainder of 
this paper, we let ${\cal W}=(1+K|y|)^{8\over{(D-2)^2a^2}}$ and ${\cal C}=
(1+\bar{K}|z|)^{4\over{(D-2)(\Delta+2)}}$ with $K$ and $\bar{K}$ respectively 
given by Eqs. (\ref{valofk}) and (\ref{kbar}) with $-$ sign.  [For the 
choice of $+$ sign, $\sigma_{DW}<0$ and $\kappa_{D-1}=0$.]  

One can determine whether gravity can be localized on the domain wall 
also by studying the linearized Einstein equations satisfied by the metric 
fluctuation $h_{\mu\nu}(x^{\rho},z)=\hat{h}^{(m)}_{\mu\nu}(x^{\rho})
{\cal C}^{-(D-2)/4}\psi_m(z)$ in the RS gauge.  When $\eta^{\mu\nu}
\partial_{\mu}\partial_{\nu}\hat{h}^{(m)}_{\mu\nu}=m^2\hat{h}^{(m)}_{\mu\nu}$, 
$\psi_m(z)$ satisfies the following Schr\"odinger-type equation 
\cite{youm1,youm3}:
\begin{equation}
-{{d^2\psi_m}\over{dz^2}}+V(z)\psi_m=m^2\psi_m,
\label{scheq}
\end{equation}
with the potential given by
\begin{equation}
V(z)={{D-2}\over{16}}\left[(D-6)\left({{{\cal C}^{\prime}}\over{\cal C}}
\right)^2+4{{{\cal C}^{\prime\prime}}\over{\cal C}}\right]=
-{{(1+\Delta)\Lambda}\over{2\Delta(1+\bar{K}|z|)^2}}+{{2\bar{K}}\over
{\Delta+2}}\delta(z).
\label{potential}
\end{equation}
Note, the potential expression in Eq. (24) of Ref. \cite{youm1} is 
related to this expression through $Q={\Delta\over{\Delta+2}}\bar{K}=
\pm\Delta\sqrt{-{\Lambda\over{2\Delta}}}$ with the sign $\pm$ being that in 
Eq. (\ref{kbar}).  [In Refs. \cite{youm1,youm4,youm2}, we considered $Q>0$ 
case, only, namely $-$ sign choice ($\sigma_{DW}>0$) for $\Delta<0$ and $+$ 
sign choice ($\sigma_{DW}<0$) for $\Delta>0$.  In this paper, we consider 
more general case, as we did in Ref. \cite{youm3}.]  Note, $V(z)$ has 
attractive [repulsive] $\delta$-function potential term when $\sigma_{DW}>0$ 
[$\sigma_{DW}<0$] for {\it any value of} $\Delta$.  So, for 
{\it any value of} $\Delta$, the graviton KK zero mode $\psi_0\sim 
{\cal C}^{(D-2)/4}$, satisfying the boundary condition $\psi^{\prime}_0(0^+)
-\psi^{\prime}_0(0^-)={{2\bar{K}}\over{\Delta+2}}\psi_0(0)$, is normalizable 
(i.e., $\int dz |\psi_0|^2<\infty$)
\footnote{Note, for the positive tension domain walls, $\bar{K}$ is positive 
[negative] for $\Delta<-2$ [$\Delta>-2$] and therefore the metric 
(\ref{cfdwsol}) is well-defined over the interval $-\infty<z<\infty$ 
[$\bar{K}^{-1}\leq z\leq -\bar{K}^{-1}$], over which the normalization 
integral is integrated}, 
as long as we choose $-$ sign in Eq. (\ref{kbar}) \cite{youm3}.  Even for 
the $\sigma_{DW}<0$ case, the graviton KK zero mode is still given by 
$\psi_0\sim {\cal C}^{(D-2)/4}$ for any values of $\Delta$ but is not 
normalizable.  Particularly, for the $\Delta<-2$ case with $\sigma_{DW}>0$, 
$V(z)$ resembles the volcano potential of the RS model \cite{rs1,rs2} and 
therefore has the same KK mode spectrum structure and the same type of 
correction to the Newtonian gravitational potential from the massive KK 
modes as the RS domain wall case, as was noted in Ref. \cite{youm1}.  
However, note that even for the $\Delta>-2$ case with $\sigma_{DW}>0$, the 
normalizable KK zero mode exists, even if $V(z)$ does not take the volcano 
potential form.  The structure of the massive KK mode spectrum for the 
$\Delta>-2$ case is discussed in Ref. \cite{youm1}.  [Note, the possible 
tachyonic modes for the $\Delta>-1$ case mentioned in Ref. \cite{youm1} do 
not exist, since the result of Ref. \cite{ceh} ensures that $m^2\geq 0$.  
For the $\Delta>0$ case, Ref. \cite{youm1} considers the case of repulsive 
$\delta$-function potential (i.e., $+$ sign choice in Eq. (\ref{kbar})), but 
the structure of the massive KK mode spectrum discussed in Ref. \cite{youm1} 
still remains the same even for the attractive $\delta$-function potential 
case.]  However, in the $\Delta>0$ case (i.e., the $\Lambda<0$ case), the 
$(D-1)$-dimensional effective action has diverging cosmological constant 
term \cite{youm3}.  It is found out \cite{youm3} that to avoid such 
divergence one has to cut off the transverse space by introducing additional 
domain wall with the fine-tuned tension between $z=0$ and the curvature 
singularity.  

To sum up, the normalizable graviton KK zero mode exists and the effective 
gravitational constant in one lower dimension is nonzero for {\it any 
values of} $\Delta$ when $\sigma_{DW}>0$, but the introduction of additional 
domain wall is required for the $\Delta>0$ case to remove diverging 
cosmological constant term in the effective action in one lower dimensions. 
And when $\sigma_{DW}\leq 0$, the graviton KK zero mode is not normalizable 
and the gravitational constant in one lower dimension is zero for {\it any 
values of} $\Delta$.  [Note, the graviton KK modes for the $\Delta=-2$ case 
was not studied by us, but was later studied in Ref. \cite{pl}.  It is shown 
there that the normalizable graviton KK zero mode exists and there is the 
continuum of the massive KK mode with mass gap for the $\Delta=-2$ case with 
$\sigma_{DW}>0$.] 

We now comment on the results of Ref. \cite{pl} which contradict the results 
of our previous papers.  The authors of Ref. \cite{pl} claim that the 
normalizable graviton KK zero mode exists only for the $\Delta\leq -2$ case 
with $\sigma_{DW}>0$, for which $\bar{K}$ (which corresponds to $k$ in Ref. 
\cite{pl}) is positive, and the introduction of additional domain wall is 
necessary for the $\Delta>-2$ case in order to trap gravity.  However, our 
careful analysis shows that gravity can be trapped even for the $\Delta>-2$ 
case, for which $k<0$ for the positive tension domain wall, and the 
introduction of additional domain wall is necessary for the $\Delta>0$ case 
not to trap gravity but to remove diverging cosmological constant term in 
the $(D-1)$-dimensional effective action.  As necessary conditions for 
trapping gravity, Ref. \cite{pl} states that the conformal factor ${\cal C}$ 
should vanish at large $|z|$ and the $\delta$-function source should have a 
positive tension so that the potential term in the Schr\"odinger equation 
can be volcano-like.  However, Ref. \cite{pl} fails to notice that ${\cal C}$ 
vanishes at finite $|z|$ (instead of at $|z|=\infty$) for the $\Delta>-2$ 
and $\sigma_{DW}>0$ case.  And as we discussed, the potential term needs not 
be volcano-like in order to support the normalizable graviton KK zero mode.

We now discuss string theory embeddings of the above-discussed domain wall 
solutions which localize gravity.  For this purpose, we begin by 
studying the uplifting of such domain walls to the dilatonic $p$-branes 
in $D^{\prime}>D$ dimensions.  The action for the dilatonic $p$-brane is 
given by
\begin{equation}
S_p={1\over{2\kappa^2_{D^{\prime}}}}\int d^{D^{\prime}}x\sqrt{-\hat{G}}
\left[{\cal R}_{\hat{G}}-{4\over{D^{\prime}-2}}(\partial\varphi)^2
-{1\over{2\cdot (p+2)!}}e^{2a_p\varphi}F^2_{p+2}\right].
\label{pbranact}
\end{equation}
The dilatonic $p$-brane solution is characterized by the parameter 
$\Delta_p=(D^{\prime}-2)a^2_p/2+2(p+1)(D^{\prime}-p-3)/(D^{\prime}-2)$.  
(See Eq. (2) of Ref. \cite{youm1} for the explicit solution with this 
convention for the action.)  Any single-charged branes in string theories are 
special cases of the dilatonic $p$-branes.  The dilatonic $p$-branes can also 
be realized from any intersecting branes (with $N$ numbers of constituents) 
in string theories by setting the charges of the constituent branes equal to 
one another and then compactifying along the relative transverse directions 
(and possibly overall transverse and longitudinal directions).  When the 
dilatonic $p$-branes are embedded in string theories in such a manner, the 
parameter $\Delta_p$ takes only the special values $4/N$ and is invariant 
under the compactification on a Ricci flat manifold involving consistent 
truncation \cite{lp}.  So, the dilatonic domain wall solutions obtained 
from (intersecting) branes in string theories through the Scherk-Schwarz 
dimensional reduction \cite{ss} on a Ricci flat manifold have $\Delta=4/N$, 
only.  However, one can have dilatonic domain wall solutions in string 
theories with different values of $\Delta$ through the compactification on 
spheres.  By compactifying the $D^{\prime}$-dimensional dilatonic 
$p$-brane (in the near horizon limit) with $\Delta_p=4/N$ on 
$S^{D^{\prime}-p-2}$, one obtains the dilatonic domain wall in 
$(p+2)$-dimensions with the dilaton coupling parameter $a$ and $\Delta$ 
given by \cite{youm1}
\begin{eqnarray}
|a|&=&{2\over p}\sqrt{{2(D^{\prime}-2)-N(p+1)(D^{\prime}-p-3)}\over
{2(D^{\prime}-p-2)-N(D^{\prime}-p-3)}},
\cr
\Delta&=&-{{4(D^{\prime}-p-3)}\over{2(D^{\prime}-p-2)-N(D^{\prime}-p-3)}}.
\label{dildelsph}
\end{eqnarray}
By applying these results, we now elaborate on various possible string 
theory embeddings of dilatonic domain walls in five dimensions.  

First, we discuss the case of the dilatonic domain walls with $\Delta=4/N$ 
($N\in{\bf Z}^+$).  To obtain such domain walls, we $(i)$ start with 
(intersecting) branes in ten or eleven dimensions (with equal constituent 
brane charges) with at least 3-dimensional (overall) longitudinal space and 
at least 1-dimensional (overall) transverse space, and then $(ii)$ 
compactify the extra (overall) longitudinal directions (if any) and all the 
relative transverse directions (if any), and perform the Scherk-Schwarz 
dimensional reduction along the extra (overall) transverse directions 
(if any).  The possible cases are as follows:
\begin{itemize}

\item $\Delta=4$ case: M5-brane; NS5-brane; D$p$-branes with $3\leq p
\leq 8$.

\item $\Delta=2$ case: $(3|{\rm M}5,{\rm M}5)$; $(3|{\rm NS}5,{\rm NS}5)$; 
$(p-1|{\rm NS}5,{\rm D}p)$ with $p=4,5,6$; $(p-2|{\rm D}p,{\rm D}p)$ 
with $p=5,6$; $(p-1|{\rm D}p,{\rm D}(p+2))$ with $p=4,5$; 
$(p|{\rm D}p,{\rm D}(p+4))$ with $p=3,4$.  

\item $\Delta=4/3$ case: M5$\perp$M5$\perp$M5 with 3-dimensional 
overall longitudinal space; D4$\perp$NS5$\perp$NS5; 
D5$\perp$D5$\perp$NS5; NS5$\perp$NS5$\perp$D5.

\end{itemize}
It is interesting to note that the spacetime metric for the 
5-dimensional domain wall solution studied in Ref. \cite{los} is 
the $\Delta=4/3$ case of the dilatonic domain wall metric.  So, the 
property of the graviton KK modes in such domain wall bulk 
background is the $\Delta=4/3$ case of what we have studied.  

Second, we discuss the string theory embeddings of the 5-dimensional 
dilatonic domain walls with $\Delta\neq 4/N$.  To obtain such domain walls, 
we $(i)$ start with (intersecting) branes in ten or eleven dimensions (with 
equal constituent brane charges) with at least 3-dimensional (overall) 
longitudinal space and at least 3-dimensional (overall) transverse space, 
$(ii)$ compactify the relative transverse directions (if any), the  
extra longitudinal directions and possibly some of overall transverse 
directions to obtain the dilatonic 3-brane in $D^{\prime}\geq 7$ 
dimensions, and then $(iii)$ compactify on $S^{D^{\prime}-5}$.  
The values of the dilaton coupling parameter $a$ and $\Delta$ of the 
resulting 5-dimensional dilatonic domain wall are the $p=3$ case of Eq. 
(\ref{dildelsph}):
\begin{eqnarray}
|a|&=&{2\over 3}\sqrt{{2(D^{\prime}-2)-4N(D^{\prime}-6)}\over
{2(D^{\prime}-5)-N(D^{\prime}-6)}},
\cr
\Delta&=&-{{4(D^{\prime}-6)}\over{2(D^{\prime}-5)-N(D^{\prime}-6)}}.
\label{dildelsph2}
\end{eqnarray}
A particularly interesting case is the non-dilatonic domain wall ($a=0$ case) 
of the RS model \cite{rs1,rs2}.  It is straightforward to check that $a$ in 
Eq. (\ref{dildelsph2}) can be zero only for the $D^{\prime}=10$ and $N=1$ 
case, i.e., the $S^5$-reduction of D3-brane.  This possibility of embedding 
the RS model was later studied in Ref. \cite{dls}.  The remaining cases of 
string theory embeddings of dilatonic domain walls with $\Delta\neq 
4/N$ are as follows:
\begin{itemize}
\item $N=1$ case:\\ 
(1) starting with M5-brane, one compactifies two of the longitudinal 
directions and then compactifies the transverse space on $T^n\times S^{4-n}$ 
($n=0,1,2$) to obtain 5-dimensional domain wall with $\Delta=-{{4(3-n)}\over
{5-n}}=-{12\over 5},-2,-{4\over 3}$;\\ 
(2) starting with NS5-brane, one compactifies two of the longitudinal 
directions and then compactifies the transverse space on $T^n\times S^{3-n}$ 
($n=0,1$) to obtain 5-dimensional domain wall with $\Delta=-{{4(2-n)}\over
{4-n}}=-2,-{4\over 3}$;\\ 
(3) starting with D$p$-brane with $3\leq p\leq 6$, one compactifies $p-3$ 
of the longitudinal directions and then compactifies the transverse space 
on $T^n\times S^{8-p-n}$ ($n\leq 6-p$) to obtain 5-dimensional domain wall 
with $\Delta=-{{4(7-p-n)}\over{9-p-n}}=-{8\over 3},-{12\over 5},-2,
-{4\over 3}$.

\item $N=2$ case:\\ 
(1) starting with $(3|{\rm M}5,{\rm M}5)$, one compactifies the relative 
transverse directions and then compactifies the overall transverse space 
on $S^2$ to obtain 5-dimensional domain wall with $\Delta=-2$;\\ 
(2) starting with $(p-1|{\rm NS}5,{\rm D}p)$ with $p=4,5$, one compactifies 
the relative transverse directions and $p-4$ of the overall transverse 
directions and then compactifies the overall transverse space on $S^2$ 
obtain 5-dimensional domain wall with $\Delta=-2$.

\end{itemize}


\begin{thebibliography} {99}
\small
\parskip=0pt plus 2pt

\bibitem{youm1} D. Youm, ``Solitons in brane worlds,'' Nucl. Phys. 
{\bf B576}, 106 (2000), hep-th/9911218.

\bibitem{rs1} L. Randall and R. Sundrum, ``A large mass hierarchy from a 
small extra dimension,'' Phys. Rev. Lett. {\bf 83}, 3370 (1999), 
hep-ph/9905221.

\bibitem{rs2} L. Randall and R. Sundrum, ``An alternative to 
compactification,'' Phys. Rev. Lett. {\bf 83} (1999) 4690, hep-th/9906064.

\bibitem{youm2} D. Youm, ``A note on solitons in brane worlds,'' 
hep-th/0001166.

\bibitem{youm3} D. Youm, ``Bulk fields in dilatonic and self-tuning flat 
domain walls,'' hep-th/0002147.

\bibitem{pl} M. Cveti\v c, H. Lu and C.N. Pope, ``Domain walls with localised 
gravity and domain-wall/QFT correspondence,'' hep-th/0007209.

\bibitem{youm4} D. Youm, ``Probing solitons in brane worlds,'' Nucl. Phys. 
{\bf B576} (2000) 123, hep-th/9912175.

\bibitem{ceh} C. Csaki, J. Erlich, T.J. Hollowood and Y. Shirman, ``Universal 
aspects of gravity localized on thick branes,'' Nucl. Phys. {\bf B581} (2000) 
309, hep-th/0001033.

\bibitem{lp} H. Lu, C.N. Pope, E. Sezgin and K.S. Stelle, ``Dilatonic 
$p$-brane solitons,'' Phys. Lett. {\bf B371} (1996) 46, hep-th/9511203; 
``Stainless super p-branes,'' Nucl. Phys. {\bf B456} (1995) 669, 
hep-th/9508042.

\bibitem{ss} J. Scherk and J.H. Schwarz, ``Spontaneous breaking of 
supersymmetry through dimensional reduction,'' Phys. Lett. {\bf B82} (1979) 60.

\bibitem{los} A. Lukas, B.A. Ovrut, K.S. Stelle and D. Waldram, 
``The universe as a domain wall,'' Phys. Rev. {\bf D59} (1999) 086001, 
hep-th/9803235.

\bibitem{dls} M.J. Duff, J.T. Liu and K.S. Stelle, ``A supersymmetric type 
IIB Randall-Sundrum realization,'' hep-th/0007120.

\end{thebibliography}
\end{document}